\documentclass[floats,floatfix,showpacs,amssymb,aps,prd,twocolumn,superscriptaddress,nolongbibliography,reprint,nofootinbib]{revtex4-2}

\usepackage{amssymb,amsmath,verbatim,mathtools,needspace,enumitem,etoolbox,graphicx,microtype,afterpage,bigints,textcomp,gensymb,tabularx,savesym,comment,nccmath,bbold,nicefrac,booktabs,footmisc,xspace,bbm,bm}

\usepackage[dvipsnames, usenames]{xcolor}
\definecolor{linkcolor}{rgb}{0.0,0.3,0.5}
\usepackage[unicode, colorlinks=true, linkcolor=linkcolor, citecolor=linkcolor, filecolor=linkcolor,urlcolor=linkcolor, pdfusetitle]{hyperref}
\usepackage[all]{hypcap}
\usepackage[T1]{fontenc}
\usepackage[utf8]{inputenc}
\usepackage{orcidlink}
\usepackage{bibunits}
\usepackage{overpic,makecell,multirow} 
\usepackage{xtab,afterpage,longtable}

\newcommand{\ssim}{\mathchar"5218\relax\,}

\makeatletter
\newcommand*{\balancecolsandclearpage}{\close@column@grid \cleardoublepage \twocolumngrid}
\makeatother

\newcolumntype{L}[1]{>{\raggedright\let\newline\\\arraybackslash\hspace{0pt}}m{#1}}
\newcolumntype{C}[1]{>{\centering\let\newline\\\arraybackslash\hspace{0pt}}m{#1}}
\newcolumntype{R}[1]{>{\raggedleft\let\newline\\\arraybackslash\hspace{0pt}}m{#1}}

\newcommand{\sm}{Supplemental Material\xspace}

\usepackage{siunitx}
\DeclareSIUnit \parsec {pc}
\DeclareSIUnit \arcsecondfull {arcsec}
\DeclareSIUnit \year{yr}
\DeclareSIUnit \day{day}
\DeclareSIUnit \hour{hr}
\DeclareSIUnit \radiant{rad}
\DeclareSIUnit \degfull{deg}
\DeclareSIUnit \erg {erg}
\DeclareSIUnit \Lsun {L_\odot}
\DeclareSIUnit \Msun {M_\odot}
\DeclareSIUnit \AstroUnit {au}
\DeclareSIUnit \erg {erg}
\usepackage{physics}
\AtBeginDocument{\RenewCommandCopy\qty\SI}
\ExplSyntaxOn
\msg_redirect_name:nnn { siunitx } { physics-pkg } { none }
\ExplSyntaxOff
\usepackage{letltxmacro}
\LetLtxMacro{\originaleqref}{\eqref}
\renewcommand{\eqref}{Eq.~\originaleqref}

\usepackage{lipsum}

\newcommand{\milan}{\affiliation{Dipartimento di Fisica ``G. Occhialini'', Universit\'a degli Studi di Milano-Bicocca, Piazza della Scienza 3, 20126 Milano, Italy}}
\newcommand{\infn}{\affiliation{INFN, Sezione di Milano-Bicocca, Piazza della Scienza 3, 20126 Milano, Italy}}
\newcommand{\geneva}{\affiliation{ D\'epartement de Physique Th\'eorique and Gravitational Wave Science Center (GWSC), Universit\'e de Gen\`eve, 24 quai Ernest Ansermet, 1211 Gen\`eve 4, Switzerland}}

\newcommand{\mrs}{\affiliation{Aix-Marseille Universit\'e, Universit\'e de Toulon, CNRS, CPT, Marseille, France}}

\begin{document}

\title{Accurate Standard Siren Cosmology with Joint Gravitational-Wave\texorpdfstring{\\}{ } and \texorpdfstring{$\gamma$}{gamma}-Ray Burst Observations}

\author{Michele Mancarella\texorpdfstring{\,}{ }\orcidlink{0000-0002-0675-508X}}
\email{mancarella@cpt.univ-mrs.fr}
\mrs \milan \infn

\author{Francesco Iacovelli\texorpdfstring{\,}{ }\orcidlink{0000-0002-4875-5862}}
\geneva

\author{Stefano Foffa\texorpdfstring{\,}{ }\orcidlink{0000-0002-4530-3051}}
\geneva

\author{Niccol\`o Muttoni\texorpdfstring{\,}{ }\orcidlink{0000-0002-4214-2344}}
\geneva

\author{Michele Maggiore\texorpdfstring{\,}{ }\orcidlink{0000-0001-7348-047X}}
\geneva


\date{\today}

\begin{abstract}
\noindent
Joint gravitational-wave (GW) and $\gamma$-ray burst (GRB) observations are among the best prospects for standard siren cosmology. However, the strong selection effect for the coincident GRB detection, which is possible only for sources with small inclination angles, induces a systematic uncertainty that is currently not accounted for. We show that this severe source of bias can be removed by inferring the \emph{a priori} unknown electromagnetic detection probability directly from multimessenger data. This leads at the same time to an unbiased measurement of the Hubble constant, to constrain the properties of GRB emission, and to accurately measure the viewing angle of each source. Our inference scheme is applicable to real data already in the small-statistics regime, a scenario that might become reality in the near future. Additionally, we introduce a novel likelihood approximant for GW events which treats the dependence on distance and inclination as exact.
\end{abstract}

\maketitle
\defaultbibliography{myrefs}
\defaultbibliographystyle{apsrev4-2}
\begin{bibunit}

\emph{Introduction---}Gravitational waves are rapidly establishing as a new pillar of concordance cosmology~\cite{Moresco:2022phi}.
A Hubble diagram can be constructed with luminosity distance measurements from gravitational--wave (GW) events 
in conjunction with redshift determinations~\cite{Schutz:1986gp}. 
The detection of a few to tens of ``bright sirens'', following GW170817~\cite{LIGOScientific:2017vwq,LIGOScientific:2017ync,LIGOScientific:2018hze}, can lead to an independent percent--level measurement of the Hubble constant $H_0$~\cite{Nissanke:2013fka,Chen:2017rfc,Feeney:2018mkj}, 
required to set the strong inconsistency between local~\cite{Riess:2021jrx} and cosmic microwave background (CMB)~\cite{Planck:2018nkj} measurements. 
With such precision, systematics will start playing a prominent role and prevent solving the Hubble tension unless properly accounted for---in particular, calibration uncertainty~\cite{Mozzon:2021wam,Huang:2022rdg}, peculiar velocities corrections~\cite{Nicolaou:2019cip,Howlett:2019mdh,Mukherjee:2019qmm}, and the electromagnetic (EM) selection effects~\cite{Chen:2020dyt}. 

The latter is the dominant source of bias when the counterpart is a short $\gamma$-ray burst (GRB)~\cite{Chen:2020dyt}, which is observed only if the angular momentum of the binary is aligned with the line of sight, i.e., at small inclination. Since inclination and luminosity distance are inherently correlated in the GW waveform, with the correlation becoming stronger for small viewing angles, neglecting this effect leads to a strong bias on $H_0$~\cite{Chen:2020dyt}. 
Knowing \emph{a priori} the GRB emission mechanism, we could impose a prior on the inclination, but the limited knowledge of these phenomena makes this impossible.
Independent measurements of the inclination from the observation of jet motions with Very Long Baseline Interferometry (VLBI) can help break the distance-inclination correlation~\cite{Mooley:2018qfh,Hotokezaka:2018dfi,Gianfagna:2023cgk,Gianfagna:2022kpw}, but these are not always guaranteed~\cite{Mastrogiovanni:2020ppa} and are themselves subject to modeling systematics~\cite{Chen:2020dyt}.
Recently, proposals for mitigating the bias in absence of direct information on the inclination and on the GRB emission were put forward. These rely on estimating the EM selection effect from a larger sample of binary neutron stars (BNSs) without counterparts~\cite{Chen:2023dgw,Gagnon-Hartman:2023soa}. 

In this Letter, we instead propose to remove the bias by determining the EM selection function solely from the joint sample of GW+GRB observations.
This scheme can be applied even with just two multimessenger events, in which case we find that neglecting the EM selection bias can lead to a $\mathcal{O}(10\%)$ shift of the posterior peak for $H_0$. 
This could happen already during currently planned runs of the LIGO-Virgo-KAGRA collaboration~\cite{Colombo:2022zzp}.
As for a larger number of events, the bias can reach the $\ssim 3\sigma$ ($\ssim 2\sigma$) level for $\mathcal{O}(50)$ [$\mathcal{O}(10)$] detections, making the correction crucial. 
Our implementation can be directly applied to real data, demonstrated with a proof of principle application to GW170817.

\medskip
\emph{Statistical formulation---}We model the problem with a hierarchical Bayesian approach~\cite{Loredo:2004nn,Mandel:2018mve,Vitale:2020aaz}. We consider a set of $N_{\rm obs}$ events with single-event detector-frame parameters $\{\pmb{\theta}_i\}_{i=1}^{N_{\rm obs}}$ and data ${\cal D} = \{{\cal D}_{\rm GW}^i, {\cal D}_{\rm EM}^i\}_{i=1}^{N_{\rm obs}}$. The EM data consist only in a redshift measurement for each event, following the identification of the host galaxy from the GRB. We instead remain agnostic to the data behind the detection of the GRB and only use the fact that such detection happened. 
For clarity, we introduce a set of boolean variables $\{{\rm det}^i_{\rm GW}, {\rm det}^i_{\rm EM}\}_{i=1}^{N_{\rm obs}}$ where ${\rm \det}^i =1$ if the event is detected and 0 otherwise. We assume that the GW catalog is obtained with a selection cut that is a known, deterministic function of the observed data only, as customary in hierarchical Bayesian analyses. 
On the other hand, the EM detections are also subject to a strong selection effect, but their selection function is not known \emph{a priori}, since we do not know the mechanism behind GRB emission. Moreover, the possibility of measuring one source parameter---the redshift---depends on another one---the inclination---which is not directly measured. This introduces a dependence of the selection function on the (also unknown) single-event parameters $\pmb{\theta}$. 
More specifically, we will model the EM detection probability, denoted by $P({\rm det}^i_{\rm EM} | \pmb{\theta}_i, \pmb{\lambda}_{\rm EM})$, as a function of the inclination angle $\iota$ and on the luminosity distance $d_L$ (for the usual scaling of the flux with $d_L^{-2}$) with universal hyperparameters $\pmb{\lambda}_{\rm EM}$ describing the GRB jet structure and the GRB detection threshold.\footnote{We assume that the viewing angle of the GRB is the same as (or a good approximation of) $\iota$ in the GW signal, and that it has a universal structure (see e.g.~\cite{Farah:2019tue,Salafia2023sjx}).}
We do not know the parameters $\pmb{\lambda}_{\rm EM}$ \emph{a priori}, but we can infer them from the GW and redshift data and, crucially, from the information that the counterpart has been detected, i.e. ${\rm det}^i_{\rm EM}=1$ for each event in the catalog.

The joint posterior on $ \{ \pmb{\lambda}, \{\pmb{\theta}_i \} \}$ can be written as
\begin{equation}\label{h_like}
\begin{split}
        p\big(&\pmb{\lambda}, \{ \pmb{\theta}_i \} | \{ {\cal D}_{\rm GW}^i\},  \{{\cal D}_{\rm EM}^i\}, \{{\rm det}_{\rm GW}^i\}, \{{\rm det}_{\rm EM}^i\} \big) \propto \\
        &\frac{\pi(\pmb{\lambda})}{P({\rm det} | \pmb{\lambda})^{N_{\rm obs}}}\times \prod_{i=1}^{N_{\rm obs}} \mathcal{L}\big( {\cal D}_{\rm GW}^i| \pmb{\theta}_i \big) \, \mathcal{L}\big({\cal D}_{\rm EM}^i| \pmb{\theta}_i , \pmb{\lambda}_{\rm c} \big) \\
        &\times p_{\rm pop}(\pmb{\theta}_i | \pmb{\lambda}) \, P({\rm det}^i_{\rm EM} | \pmb{\theta}_i, \pmb{\lambda}_{\rm EM})\, .
    \end{split}
\end{equation}

A detailed derivation and discussion of this posterior is 
provided in Supplemental Material~\cite{nn:supplementA}.\footnote{See also~\cite{Mould:2023eca} for a similar discussion in the context of GW data only.} 
In \eqref{h_like}, $p_{\rm pop}(\pmb{\theta}_i | \pmb{\lambda})$ is the probability of a source to have parameters $\pmb{\theta}_i$ given hyperparameters $\pmb{\lambda}$, $\pmb{\lambda}_{\rm c} = \{H_0, \Omega_{\rm m,0}\}$ are the cosmological parameters, $\mathcal{L}\big( {\cal D}_{\rm GW}^i| \pmb{\theta}_i \big)$ and $ \mathcal{L}\big({\cal D}_{\rm EM}^i| \pmb{\theta}_i , \pmb{\lambda}_{\rm c} \big)$ are the GW and EM likelihoods, $\pi(\pmb{\lambda})$ is the prior on hyperparameters, and $P(\rm det | \pmb{\lambda})$ denotes the fraction of detectable events~\cite{Mandel:2018mve}. This term contains both the GW and EM selection effects, with the latter given by $P({\rm det}^i_{\rm EM} | \pmb{\theta}_i, \pmb{\lambda}_{\rm EM})$.

The key difference between \eqref{h_like} and the standard hierarchical Bayesian posterior in GW population studies is the presence of the term $P({\rm det}^i_{\rm EM} | \pmb{\theta}_i, \pmb{\lambda}_{\rm EM})$ at the numerator. A similar contribution is not present for GW data because detectability is a property of data only, which allows us to write $\mathcal{L}({\rm det}_{\rm GW}^i, {\cal D}_{\rm GW}^i | \pmb{\theta}_i ) = P({\rm det}_{\rm GW}^i | {\cal D}_{\rm GW}^i) \mathcal{L}({\cal D}_{\rm GW}^i | \pmb{\theta}_i ) = \mathcal{L}({\cal D}_{\rm GW}^i | \pmb{\theta}_i ) $, where the first equality follows from the composition of probabilities and the second from the fact that for detected events $P({\rm det}_{\rm GW}^i | {\cal D}_{\rm GW}^i)=1$.\footnote{Note, however, that the GW detection probability, though absent from the numerator of \eqref{h_like}, is present at the denominator in the term $P(\rm det | \pmb{\lambda})$ and needs to be properly computed.} A similar simplification does not hold for the EM likelihood due to the discussed dependence of the detection probability on $\pmb{\theta}_i$ and $\pmb{\lambda}_{\rm EM}$.
The interpretation of the detection probability $P({\rm det}^i_{\rm EM} | \pmb{\theta}_i, \pmb{\lambda}_{\rm EM})$ in \eqref{h_like} is that this term will reshape the GW likelihood essentially acting as a modification of the prior~\cite{Essick:2023upv,Mould:2023eca} carrying information about the fact that the detection of the GRB emission preferentially selects small viewing angles. This breaks the distance-inclination correlation of each individual GW event likelihood leading to an unbiased estimate of $H_0$ (as well as of other cosmological parameters) and to the reconstruction of the maximum viewing angle.

\medskip
\emph{A new GW likelihood approximant---}The GW data are usually given, for each event $i$, in terms of samples from the posterior probability $p\big(  \pmb{\theta}_i | {\cal D}_{\rm GW}^i\big) \propto \mathcal{L}\big( {\cal D}_{\rm GW}^i | \pmb{\theta}_i \big) \pi({\pmb{\theta}_i})$ where $\pi({\pmb{\theta}_i})$ is the prior used to analyze the data.
However, full Bayesian simulations are quite expensive, especially for BNSs, and are often circumvented by resorting to the Fisher Information Matrix (FIM) approximation, i.e., a Gaussian approximation of the likelihood obtained by an expansion of the signal around the true parameters' values~\cite{Vallisneri:2007ev}. This, however, fails precisely in the corner of the parameter space which is relevant for GRBs, i.e. at small or vanishing inclination angle~\cite{Iacovelli:2022bbs,Dupletsa:2024gfl}, regardless of possible regularizations  
or of the use of priors~\cite{Vallisneri:2007ev,Dupletsa:2022scg,Iacovelli:2022bbs,Dupletsa:2024gfl}.

For this Letter, we develop a new likelihood approximant that treats the likelihood in the subspace $\{d_L, \iota\}$ as exact, while resorting to the FIM in the rest of the parameter space. This extends existing approaches, e.g.,~\cite{Chen:2018omi,Chassande-Mottin:2019nnz,deSouza:2023gjv,Li:2024iua}. 
The result can be expressed in terms of a simple combination of FIM elements, computable with open--source packages~\cite{Borhanian:2020ypi,Dupletsa:2022scg,Iacovelli:2022mbg}.
We also introduce a second extension of the FIM approximation, in order to properly encode the effects of noise fluctuations~\cite{Essick:2023upv}.
This consists of adopting an expansion around the point maximizing the likelihood in presence of noise, which we obtain numerically from an explicit realization of the noise in the frequency domain. 
The procedure is described in detail in Supplemental Material~\cite{n:supplementB}. 
\begin{figure}[t]
    \centering
    \includegraphics[width=.5\textwidth]{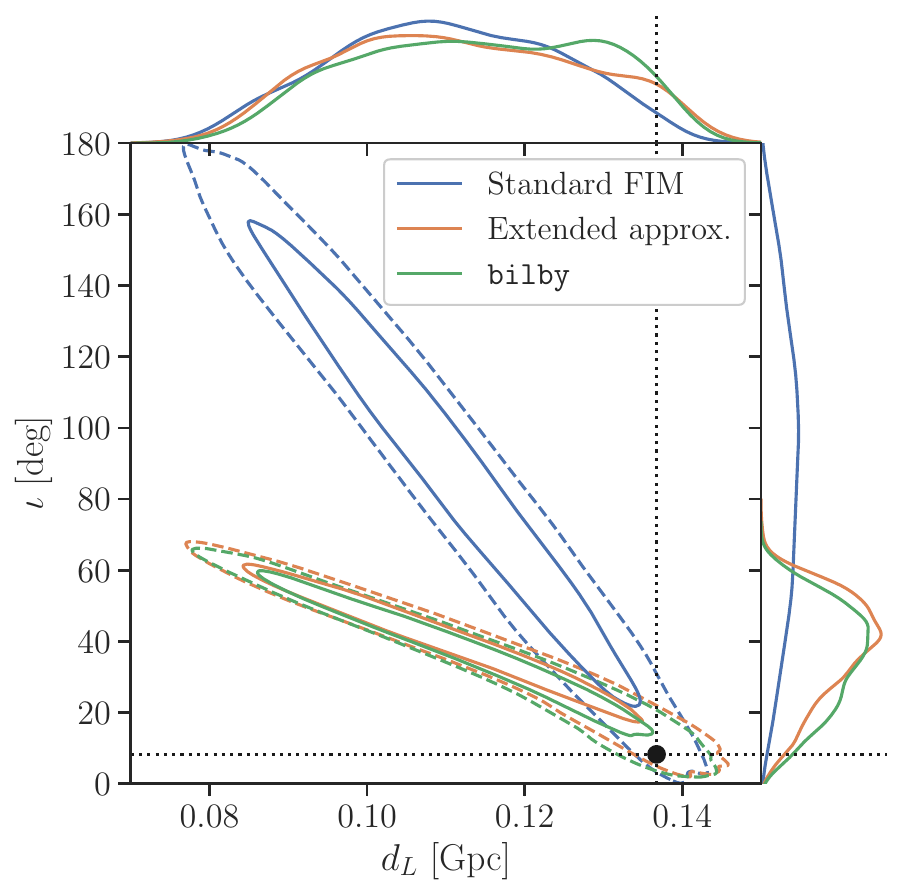}
    \caption{Posterior probability in the distance-inclination plane for the highest SNR source in the simulations used for the results of this Letter. 
    Orange contours are obtained using our new likelihood approximant which encodes the exact dependence of the likelihood on $d_L$ and $\iota$. 
    Blue contours show for comparison the Fisher matrix approximation, which is inaccurate for sources at low inclination.
    Green contours are obtained from a Bayesian parameter estimation with \texttt{PARALLEL BILBY}.}
    \label{fig:dLiota}
\end{figure}

\autoref{fig:dLiota} shows posterior contours in the distance-inclination plane for the highest SNR source in our simulation, obtained with our extended likelihood approximant (orange), compared to the FIM (blue), and, as proof of the validity of the extended approximant, to a full Bayesian simulation with \texttt{PARALLEL BILBY}~\cite{Ashton:2018jfp,Smith:2019ucc} (green). We see that the FIM completely mismodels the shape of the likelihood, while the agreement between our approximant and the full likelihood is excellent. 
The biased reconstruction of the marginal distance posterior in \autoref{fig:dLiota} is a consequence of the small inclination of the source and is therefore intrinsic to any multimessenger event with a GRB, and the origin of the systematic uncertainty on $H_0$.\footnote{The bias is further exacerbated by the use of a prior on inclination flat in $\cos \iota$, following the assumptions that inclinations have a uniform distribution on the sphere.}

\begin{figure}[t]
    \centering
    \includegraphics[width=.5\textwidth]{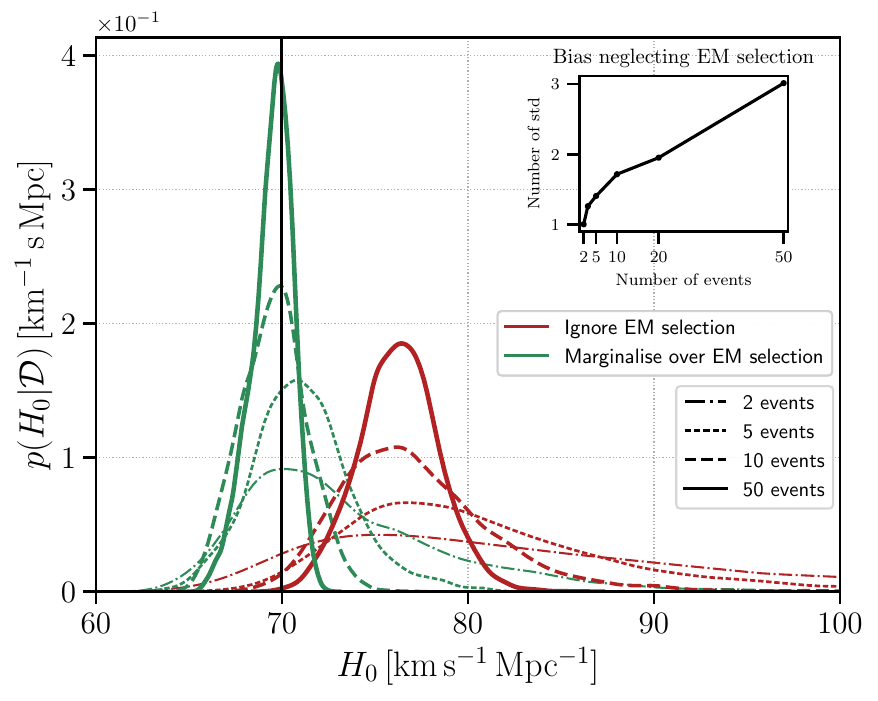}
    \caption{ Marginal posterior probability on $H_0$, without (red) and with (green) the inclusion of the EM detection probability. Different line styles correspond to a different number of events analyzed, as in the legend. Inset: bias on the measurement when neglecting the EM selection probability.} 
    \label{fig:Homarginal}
\end{figure}

This simulation scheme is indispensable for studies of third-generation (3G) detectors, for which an efficient and scalable Bayesian inference sampling tool is still missing (but see~\cite{Nitz:2021pbr,Smith:2021bqc,Wong2023lgb,Wouters:2024oxj} for recent progress), while detection rates are expected to be large~\cite{Ronchini:2022gwk,Iacovelli:2022bbs}. 

\medskip
\emph{Case study---}We now illustrate the method on a simple yet realistic example for the current generation of GW and EM experiments, namely, a LIGO-Virgo-KAGRA-LIGO India (LVKI) network~\cite{LIGOScientific2014pky,VIRGO2014yos,Aso2013eba,LIGOI,LVKIliving} and Fermi-$\gamma$-ray burst monitor (GBM)~\cite{Meegan:2009qu}.

We simulate a population of BNSs with a Madau-Dickinson profile for the merger rate~\cite{Madau:2014bja}, a Gaussian mass distribution, and inclinations uniformly distributed on the sphere.
We assume an LVKI network at design sensitivity with a $100\%$ duty cycle and threshold on the observed SNR $\rho_{\rm obs} \geq 12$~\cite{LVKIliving}.
For the EM detection model, we assume that the GRB jet has a Gaussian-shaped profile with amplitude $A_0$ and half-opening $\Theta_{\rm c}$, compatible with the multimessenger analysis of GW170817~\cite{Lyman:2018qjg,Troja:2018ruz,Resmi:2018wuc}. 
We consider an event detected if the flux exceeds the Fermi-GBM-like sensitivity of $F_{\rm th} = \SI{2e-7}{\erg\per\square\centi\meter\per\second}$~\cite{Meegan:2009qu}.
This model predicts a maximum viewing angle (defined as  $\Theta={\rm min}\{\iota,\, 180^{\circ} - \iota\}$)
of order $\Theta_{\rm max}\sim11^{\circ}$ (\SI{0.2}{\radiant}) in the redshift range of interest for our sample (see \autoref{fig:theta_vs_z} below). 
Our simulated dataset consists of a set of samples from the posterior $p(\pmb{\theta}_i | {\cal D}_{\rm GW}^i) \propto \mathcal{L}\big( {\cal D}_{\rm GW}^i| \pmb{\theta}_i \big) / \pi(\pmb{\theta}_i)$ obtained with the likelihood approximant introduced before, and a redshift measurement with uncertainty $\sigma_z = 10^{-3}$ for each GW event. 
We sample \eqref{h_like} in the high-dimensional space of $\pmb{\lambda}, \{ \pmb{\theta}_i \}$ using \texttt{PYMC}~\cite{pymc2023} and the Hamiltonian Monte-Carlo-based scheme NUTS~\cite{Brooks_2011,JMLR:v15:hoffman14a}. 
The hyperparameters $\pmb{\lambda}$ include cosmology, mass, and redshift distributions, together with the parameters $\pmb{\lambda}_{\rm EM}= \{ A_0,\, \Theta_{\rm c}, F_{\rm th} \}$ describing the EM detection probability. 
A thorough description of the population model, of the mock catalog and of the inference scheme is provided in Supplemental Material~\cite{note:supplementC}.

\autoref{fig:Homarginal} shows the effect of neglecting (red lines) and inferring (green lines) the EM detection probability on the posterior probability on $H_0$, marginalized over all the remaining hyperparameters and individual event parameters, for 2, 5, 10 and 50 events. When neglecting the EM detection probability a non-negligible ($\ssim 7-10\%$) shift of the peak is already visible for two events, remaining stable as the number of detections increases. The simultaneous fit of the EM selection function can remove the bias as far as $\geq 2$ events are included in the analysis. 
Of course, for low statistics, the bias is partially compensated by the large statistical uncertainty. The inset in \autoref{fig:Homarginal} quantifies the bias as the ratio between the shift of the posterior peak from the injected value and the half $68.3 \%$ highest density interval, which we denote by $\sigma$ being this quantity the standard deviation in the Gaussian case. This reaches the $\ssim 2\sigma$ level between 10 and 20 events and $\ssim 3\sigma$ for 50 events. We conclude that while for low detection rates the shift of the peak could be somewhat compatible with a noise fluctuation, only the inclusion of the EM selection effect can ensure a robust measurement. For a larger number of detections, as might be expected in a few years of O4--O5 observations~\cite{Colombo2022} and should be expected for 3G detectors~\cite{Ronchini:2022gwk}, the bias is statistically significant and must be corrected.
We have verified that the unbiased estimate of $H_0$ is not spoiled even in presence of a mismatch between the generative model of the EM detection probability and the one used for inference, for which we also tried a sigmoid model and a polynomial one. The explicit functional forms are provided in the Supplemental Material~\cite{note:supplementC}. This ensures that the main result of this Letter, that is, the elimination of the bias on $H_0$, is independent on the details of the GRB jet structure assumed in the analysis.

\begin{figure}[tbp]
    \centering
    \includegraphics[width=.5\textwidth]{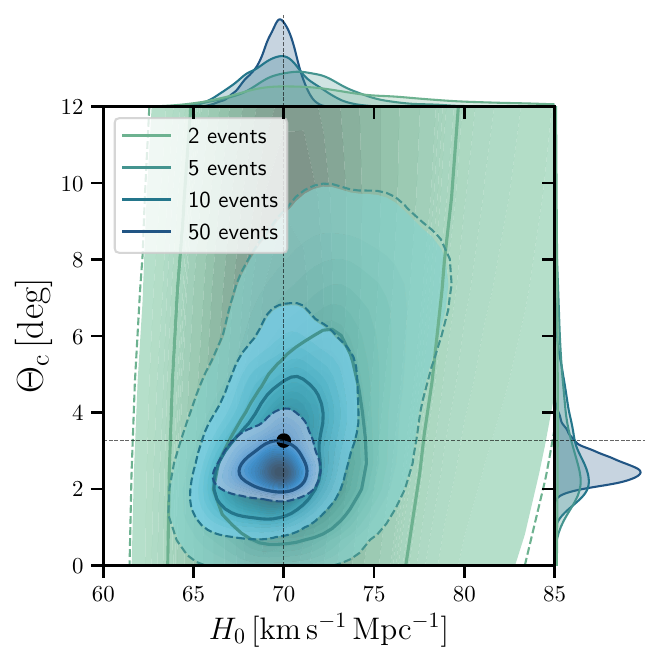}
    \caption{ Joint constraint on $H_0$ and the half-width of the jet $\Theta_{\rm c}$. Dashed (solid) lines indicate the $95\%$ ($68\%$) confidence level.}
    \label{fig:H0iota}
\end{figure}

\autoref{fig:H0iota} shows the correlation between $H_0$ and the half--width of the jet $\Theta_{\rm c}$, when the EM detection probability is inferred from the data. In particular, for $\gtrsim 10$ events, large values of $\Theta_{\rm c}$ ($\gtrsim 5^{\circ}$) are excluded. This shows that adopting the strategy of putting a prior on the inclination of single events without inferring it from data can lead to a biased estimate of $H_0$.

Furthermore, our inference scheme allows us to reconstruct the posterior probabilities of individual source parameters for each event, properly reweighted by the population prior and the EM detection probability.
Remarkably, an accurate measurement of the viewing angle for each event is obtained just from the deconvolution of the latter, even in absence of a direct measurement of this angle in the EM (see~\cite{Farah:2019tue} for a similar approach).
These unbiased measurements are shown in \autoref{fig:theta_vs_z} as a function of redshift for the case of ten events as green dots,\footnote{We find that 50 events at O5 sensitivity give only a marginal improvement due to the small horizon of the GW detectors.} and recover the injected values (black stars) within $\ssim 1 \sigma$ with good accuracy. 
The inferred values are below the reconstructed maximum viewing angle $\Theta_{\rm max}$ at any redshift (green band).
In contrast, the measurement from GW data only (blue points) is largely inaccurate, as expected (see \autoref{fig:dLiota}). 

Finally, to provide an idea of the information that we can extract with this technique from the only event with counterpart detected so far, and to show that our pipeline can be straightforwardly applied to actual detections, the violin point in yellow in \autoref{fig:theta_vs_z} shows the result obtained with the analysis of GW170817 and its counterpart, assuming a Gaussian-shaped profile for the jet.\footnote{We use the ``high spin'' posterior samples available on \href{https://gwosc.org}{GWOSC} \cite{LIGOScientific:2019lzm}, the redshift given in~\cite{LIGOScientific:2017adf} and compute the detection probability with an injection set generated with O2 sensitivity and an optimal SNR cut $\rho_{\rm opt}\geq10$~\cite{LIGOScientific:2021psn}. We find that the posterior probability on $H_0$ is perfectly compatible with the result of~\cite{LIGOScientific:2017adf} even when including the EM detection probability, as expected with a sample of 1 event only and in line with the results of our simulations.}

\begin{figure}[tbp]
    \centering
    \includegraphics[width=.5\textwidth]{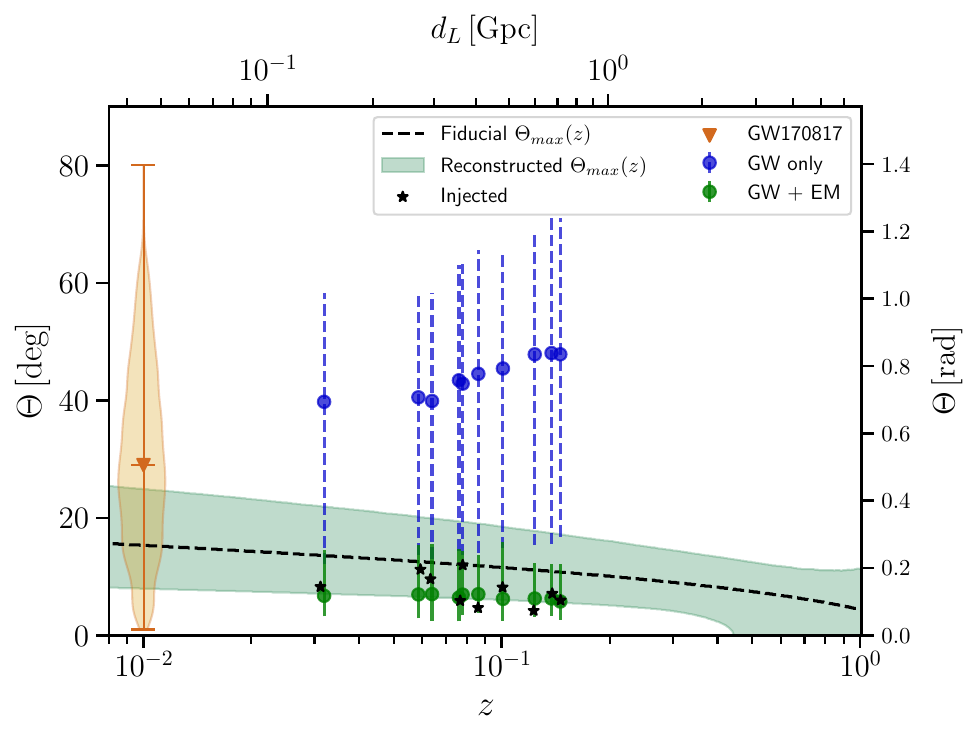}
    \caption{Reconstruction of the maximum viewing angle of the GRB as a function of redshift from ten simulated multimessenger events. The fiducial model is shown as a black, dashed line. Dots with error bars indicate the measurements without (blue) and with (green) the inclusion of the EM detection probability. Black stars denote the injected values. The yellow violin plot shows, for comparison, the result for the viewing angle of GW170817 obtained with the method proposed in this Letter.}
    \label{fig:theta_vs_z}
\end{figure}

\medskip
\emph{Conclusions and discussion---}The bias arising from EM selection effects can be a major limitation for standard siren cosmology already in the short term. In this Letter, we provided a new method to properly treat it. 
We can summarize the main novelties as follows:

\begin{itemize}

    \item \emph{Unbiased inference of $H_0$ in the short term.} 
    It is possible to accurately measure $H_0$ by inferring the EM detection probability from multimessenger data only and without the need of modeling explicitly the GRB detector, starting from as few as two sources, which could become a concrete scenario in the near future. In this case, the inference scheme introduced here would be directly applicable, demonstrated by applying our pipeline to GW170817. 
    \item \emph{Determination of the GRB emission profile and individual viewing angles.} 
    The proposed inference scheme leads to a reconstruction of the GRB emission profile as well as to an accurate and precise measurement of the inclination of each event. The same can be used in conjunction with a prior on $H_0$ (from either CMB or supernovae) to determine the GRB structure only from GW data and redshift measurements (see~\cite{Hayes2019hso,Hayes2023zxm,Salafia2023sjx}). In this scenario, different models of the emission can be tested against the data, e.g. by comparing the respective Bayes factors~\cite{Hayes2019hso,Hayes2023zxm}. On the other hand, an unbiased estimate of $H_0$ is obtained even in presence of a mismatch between the true, unknown emission profile and the model used to analyze the data, making the Hubble constant measurement robust to modeling systematics.

\end{itemize}

These results open other possibilities: 

\begin{itemize}

    \item \emph{More accurate modeling of the EM emission.} 
    More general models of the EM selection function can be straightforwardly included, such as a limited redshift range for the EM detections, a dependence of the EM emission on the masses, a selection of GW sources based on their sky localization, and the field of view of the EM instrument. We plan to address these aspects in future work. 

    \item \emph{Model--independent approaches.} In this Letter, for simplicity, the functional dependence of the flux from inclination used for inference was the same used to generate the EM data. Importantly, as already discussed, this assumption is not essential to the method. One could adopt a more agnostic approach and model the EM detection probability as a flexible function (e.g. a polynomial or a Gaussian process) in a purely data-driven perspective.

    \item \emph{Independent inclination measurements.} We assumed that independent inclination measurements (e.g., from VLBI) are not present to stress the generality of the method. Nonetheless, such measurements can be straightforwardly included by adding the corresponding likelihood in \eqref{h_like}.  
    In this case, however, additional sources of bias might be present~\cite{Chen:2020dyt}, which can be cured with the introduction of additional parameters~\cite{Salvarese:2024jpq}.

    \item \emph{Population properties and perspectives for 3G detectors.} GW and EM observatories of next generation are under active development~\cite{Maggiore:2019uih,Evans:2021gyd,Branchesi:2023mws}. In addition to $H_0$, our inference scheme includes the reconstruction of the mass and redshift distributions of the BNS population, as well as the full distance-redshift relation, including the matter energy density and the dark energy equation of state. It can also be straightforwardly extended to test the effect of modified GW propagation~\cite{Belgacem:2017ihm,Belgacem:2018lbp}. With 3G sensitivities, we will be able to precisely determine all of those. We will provide forecasts in future work.
    
\end{itemize}

With more multimessenger events in the near future and major new experiments planned, we hope that this Letter will contribute to advancing toward accurate and precise standard siren cosmology both in the short and long term.\\[1.cm]

\emph{Acknowledgments---}We thank Simone Mastrogiovanni for LVK internal review; Om Salafia, Matthew Mould, Davide Gerosa, and Hsin-Yu Chen for useful comments on the draft; Marica Branchesi and Samuele Ronchini for useful exchanges in the initial stage of the work; Massimo Dotti, Costantino Pacilio, Arianna Renzini, Pippa Cole, and Alberto Colombo for discussions.
The work of M.Mancarella received support from the French government under the France 2030 investment plan, as part of the Initiative d'Excellence d'Aix-Marseille Universit\'e -- A*MIDEX AMX-22-CEI-02.
M.Mancarella is supported by European Union's H2020 ERC Starting Grant No. 945155--GWmining and Cariplo Foundation Grant No. 2021-0555.
F.I, S.F., M.Maggiore and  N.M.  are supported by  the  Swiss National Science Foundation, Grant No. 200020$\_$191957, and  by the SwissMap National Center for Competence in Research. 
This work made use of the clusters Yggdrasil and Baobab at the University of Geneva. 

\emph{Data availability---}\texttt{GWFAST} is publicly available at \href{https://github.com/cosmostatgw/gwfast}{github.com/cosmostatgw/gwfast}. \texttt{PYMC} is availabe at \href{https://www.pymc.io/welcome.html}{pymc.io}.

This document has LIGO DCC number LIGO-P2400167.

\putbib
\end{bibunit}

\begin{bibunit}
\setcounter{equation}{0}
\setcounter{figure}{0}
\setcounter{table}{0}
\clearpage \pagebreak

\section*{\sm}

\section{Derivation of the hierarchical posterior}\label{app:hierarchical_posterior}

We consider a set of events with single-event parameters $\{\pmb{\theta}_i\}_{i=1}^{N_{\rm obs}}$ and data $\{ {\cal D}\}_{i=1}^{N_{\rm obs}}$.  We also introduce a set of boolean variables $\{{\rm det}^i\}_{i=1}^{N_{\rm obs}}$ where ${\rm \det}^i =1$ if the event is detected and 0 otherwise.
As common in GW population studies, the hierarchical model is an inhomogeneous Poisson process~\cite{Loredo:2004nn,Mandel:2018mve}.
The expected number of sources is defined as $N_{\rm exp}(\pmb{\lambda}) = N \, P(\rm det | \pmb{\lambda}) $ with
\begin{equation}
     P({\rm det} | \pmb{\lambda})  = \int \dd\pmb{\theta} \dd{\cal D} \, p_{\rm pop}(\pmb{\theta} |\pmb{\lambda} ) \mathcal{L}({\cal D}|\pmb{\theta}) \, P({\rm det} | {\cal D}, \pmb{\theta}, \pmb{\lambda})\, .
\end{equation}
In the literature, $P(\rm det | \pmb{\lambda})$ is often denoted as $\alpha({\pmb{\lambda}})$,\footnote{The notation $\beta({\pmb{\lambda}})$ or $\sigma({\pmb{\lambda}})$ is also adopted.} and represents the fraction of detectable events given hyperparameters ${\pmb{\lambda}}$, while usually the integral $\int \dd {\cal D} \mathcal{L}({\cal D}|\pmb{\theta}) \, P({\rm det} | {\cal D}, \pmb{\theta}, \pmb{\lambda}) = \int_{\raisebox{2pt}{$\scriptstyle f({\cal D})\geq 0$}} \dd {\cal D} \mathcal{L}({\cal D}|\pmb{\theta}) $ is denoted as $P_{\rm det}(\pmb{\theta})$ [or more correctly $P({\rm det} | \pmb{\theta})$], under the assumption that $P({\rm det}_i | {\cal D}, \pmb{\theta}, \pmb{\lambda}) = P({\rm det} | {\cal D})$ and that $P({\rm det} | {\cal D})$ is an indicator function equal to 1 if $f({\cal D})\geq0$ and 0 otherwise, see below. Concretely, in GW population studies one often assumes $f({\cal D}) = \rho_{\rm obs} - \rho_{\rm th}$ where $\rho_{\rm obs}$ is the observed SNR in the detector and $\rho_{\rm th}$ a given threshold that defines a detection.

Including explicitly the variables $\{{\rm det}^i\}_{i=1}^{N_{\rm obs}}$, the hierarchical likelihood can be written as~\cite{Mandel:2018mve,Essick:2023upv}
\begin{equation}\label{generalLik}
\begin{split}
    p\big(  \{ \pmb{\theta}_i \}, \{ {\cal D}_i\}, \{{\rm det}_i\} |\pmb{\lambda}  \big) = &\,N^{N_{\rm obs}} e^{-N P(\rm det | \pmb{\lambda})} \\
    &\,\times \prod_{i=1}^{N_{\rm obs}} p({\rm det}_i, {\cal D}_i, \pmb{\theta}_i | \pmb{\lambda})\, ,
\end{split}
\end{equation}
where in full generality 
\begin{equation}
    \begin{split}
        p({\rm det}_i, {\cal D}_i, \pmb{\theta}_i | \pmb{\lambda} ) &= \mathcal{L}({\rm det}_i, {\cal D}_i | \pmb{\theta}_i ) p_{\rm pop}(\pmb{\theta}_i |\pmb{\lambda} ) \\
        &= P({\rm det}_i | {\cal D}_i, \pmb{\theta}_i, \pmb{\lambda})  \mathcal{L}({\cal D}_i | \pmb{\theta}_i )  p_{\rm pop}(\pmb{\theta}_i |\pmb{\lambda})\,.
    \end{split}
\end{equation}
At this point, the usual assumption in GW astronomy is that detectability is a function of the data only, and not of the (true, unknown) source parameters $\pmb{\theta}_i$, nor of the hyperparameters $\pmb{\lambda}$. This implies $P({\rm det}_i | {\cal D}_i, \pmb{\theta}_i, \pmb{\lambda}) = P({\rm det}_i | {\cal D}_i)$. Finally, assuming that detection is a deterministic function of the data, one has $P({\rm det}_i | {\cal D}_i)=1 \, \forall i$ since each event in the catalog has obviously been detected.\footnote{It can be shown that the likelihood is independent of this term even in the more general case in which the detection is a probabilistic function of the data provided that it is independent of the true source parameters $\pmb{\theta}_i$~\cite{Essick:2023upv}.} This is what we assume for GW data also in this work---an event is detected if the \emph{observed} SNR exceeds a given threshold.

However, we might be in a situation where detectability depends on some (unknown) source parameters that are not measured by the experiment. In this case, we have no data available for all the parameters determining detectability, but we know that a detection happened, and we can include this information. In particular, we assume that detectability has an a--priori unknown dependence on some of the (equally unknown) source parameters $\pmb{\theta}_i$ through a subset of hyperparameters $\tilde{\pmb{\lambda}}$ that we wish to determine from data. In this case, $P({\rm det}_i | {\cal D}_i, \pmb{\theta}_i, \tilde{\pmb{\lambda}})$ is not equal to one for all $\pmb{\theta}_i$. This term will appear in the likelihood and the parameters $\tilde{\pmb{\lambda}}$ can be determined from the data. This is the case for the EM data in this work, as we now detail.

Turning to the specific case of interest for this paper, we assume that the likelihood is composed by GW ad EM data with independent noise, $\mathcal{L}({\cal D} | \pmb{\theta}, z ) = \mathcal{L}\big( {\cal D}_{\rm GW}| \pmb{\theta} \big) \, \mathcal{L}\big( z_{\rm EM}^{\rm obs}| z \big)$. We denote by $\{ {\cal D}_{\rm GW}^i,  {\cal D}_{\rm EM}^i\}_{i=1}^{N_{\rm obs}}$ the two datasets, with ${\cal D}_{\rm EM} = \{ z_{\rm EM}^{\rm obs} \}$. The parameters $\pmb{\theta}$ refer to detector--frame parameters of the GW waveform while $z$ denotes the true redshift. The GW likelihood is discussed in Sec.~\ref{app:lik}. As anticipated, the GW measurement is modeled as a deterministic function of the data.

Turning to the EM sector, crucially, in this work we assume that we are not able to obtain a measurement of all the variables determining the EM detections. 
Explicitly, for the problem under consideration in this paper, the EM data consist only in a redshift measurement, following from the identification of the host galaxy from the GRB emission.\footnote{Notice that, even if we assume an EM likelihood for the redshift only, crucially the counterpart detection provides also a very accurate measurement of the sky position of the source and of the GPS time of the event. This information in our setting is already incorporated in the GW likelihood, which we compute keeping these parameters fixed.} We expect that the detection of the source can happen only under certain conditions on its inclination (namely, the inclination has to be small enough in order to observe a relativistic jet), but we remain agnostic about the data from the GRB observatory (see~\cite{Salafia2023sjx} for an explicit modeling), and assume that we are just given an observed redshift $ z_{\rm EM}^{\rm obs}$ (for which we will assume a Gaussian likelihood with standard deviation $\sigma_{z}$). 

We then assume that the EM detection has a dependence on the source parameters and the hyperparameters $\pmb{\lambda}_{\rm EM}$ describing the GRB jet structure. Hence, $P({\rm det} | {\cal D}, \pmb{\theta}, \tilde{\pmb{\lambda}}) = P({\rm det}_{\rm GW} | {\cal D}_{\rm GW}) 
P({\rm det}_{\rm EM} | \pmb{\theta}, \pmb{\lambda}_{\rm EM}) = P({\rm det}_{\rm EM} | \pmb{\theta}, \pmb{\lambda}_{\rm EM})$. 

Finally, the population prior is
\begin{equation}\label{ppop}
    p_{\rm pop}(\pmb{\theta} | \pmb{\lambda}) \propto {\frac{\psi(z, \pmb{\lambda}_{\rm z})}{1+z} \frac{\dd V_c}{\dd z}\;  \frac{p( m_1, m_2 | \pmb{\lambda}_{\rm m}) \, } { \frac{\partial{d_L}}{\partial z}(z, \pmb{\lambda}_{\rm c} ) \, (1+z)^2} }\Bigg|_{\raisebox{6pt}{$\scriptstyle\substack{m_{i}=\frac{m_{i}^{\rm det}}{1+z(d_L, \pmb{\lambda}_{\rm c})}\\z=z(d_L, \pmb{\lambda}_{\rm c})}$}} \,,
\end{equation}
where $p( m_1, m_2 | \pmb{\lambda}_{\rm m})$ is the source--rame mass distribution with hyperparameters $\pmb{\lambda}_{\rm m}$,  $\psi(z, \pmb{\lambda}_{\rm z})$ is the source--rame merger rate distribution with hyperparameters $\pmb{\lambda}_{\rm z}$, $\dd V_c/\dd z$ is the differential comoving volume, the factor $1/(1+z)$ takes into account time dilation from source to detector frame, and the factor $(1+z)^2\partial{d_L}/\partial z(z, \pmb{\lambda}_{\rm c})$ at the denominator comes from conversion of the prior from detector- to source-frame variables. Explicitly, the distance-redshift relation in a flat $\Lambda$CDM Universe is given by 
\begin{equation}\label{dLofz}
    d_L(z, \pmb{\lambda}_{\rm c} ) =\frac{ c(1+z)}{H_0} \int_0^z \frac{\dd z'}{\sqrt{\Omega_{\rm m,0}(1+z')^3+1-\Omega_{\rm m,0}}}\,.
\end{equation}
The cosmological hyperparameters are denoted by $\pmb{\lambda}_{\rm c} = \{H_0, \Omega_{\rm m,0}\}$.
The full set of hyperparameters is defined by $\pmb{\lambda} = \{\pmb{\lambda}_{\rm c}, \pmb{\lambda}_{\rm m}, \pmb{\lambda}_{\rm z}, \pmb{\lambda}_{\rm EM}  \}$.

We will work in detector-frame; hence, in \eqref{ppop} the source-frame quantities are computed from the detector-frame ones with a dependence on the cosmology. In particular, the redshift can always be re-expressed as a function of the luminosity distance and of the cosmological parameters $\pmb{\lambda}_{\rm c}$~\cite{Finke:2021aom}; we will assume to have performed this change of variables and denote the  EM likelihood as $\mathcal{L}\big({\cal D}_{\rm EM}^i| \pmb{\theta}_i , \pmb{\lambda}_{\rm c} \big) = \mathcal{L}\big(z_{{\rm EM}, i}^{\rm obs}| z(d_{L,i}, \pmb{\lambda}_{\rm c}) \big)$.

Putting everything together in \eqref{generalLik}, adding a prior $\pi(\pmb{\lambda})$ on the hyperparameters, and marginalizing over the overall number of expected events $N$ with a prior $\propto 1/N$~\cite{Mandel:2018mve}, we arrive at the following hierarchical likelihood: 
\begin{equation}\label{h_like_supp}
\begin{split}
        p\big(&\pmb{\lambda}, \{ \pmb{\theta}_i \} | \{ {\cal D}_{\rm GW}^i\},  \{{\cal D}_{\rm EM}^i\}, \{{\rm det}_{\rm GW}^i\}, \{{\rm det}_{\rm EM}^i\} \big) \propto \\
        &\frac{\pi(\pmb{\lambda})}{P({\rm det} | \pmb{\lambda})^{N_{\rm obs}}}\times \prod_{i=1}^{N_{\rm obs}} \mathcal{L}\big( {\cal D}_{\rm GW}^i| \pmb{\theta}_i \big) \, \mathcal{L}\big({\cal D}_{\rm EM}^i| \pmb{\theta}_i , \pmb{\lambda}_{\rm c} \big) \\
        &\times p_{\rm pop}(\pmb{\theta}_i | \pmb{\lambda}) \, P({\rm det}^i_{\rm EM} | \pmb{\theta}_i, \pmb{\lambda}_{\rm EM})\, .
    \end{split}
\end{equation}
\medskip

Interestingly, we can also make an explicit link to the modeling of the likelihood of the GRB experiment~\cite{Salafia2023sjx}.\footnote{We are grateful to Om Salafia for discussions on this point.} In this case, one would observe a flux with the observation described by a likelihood of the form $\mathcal{L}(F^{\rm obs} | F)$. Each term in the product at the numerator of \eqref{h_like_supp} would then be of the form
\begin{equation}
\begin{split}
    &\mathcal{L}\big( {\cal D}_{\rm GW}^i| \pmb{\theta}_i \big) \, \mathcal{L}\big(z_{{\rm EM}, i}^{\rm obs}| z(d_{L,i}, \pmb{\lambda}_{\rm c})\big) \, \mathcal{L}(F^{\rm obs}_i | F_i) \, \\
    &\ \times p_{\rm pop}(\pmb{\theta}_i | \pmb{\lambda}) \, p_{\rm pop}(F| \pmb{\theta}_i, \pmb{\lambda}_{\rm jet}) \, , 
\end{split}
\end{equation}
where $\pmb{\lambda}_{\rm jet}$ are the parameters describing the GRB jet structure (e.g. $\{ A_0, \Theta_{\rm c} \}$ in this paper).
Being agnostic on the GRB data is equivalent to marginalize the likelihood over the observed flux $F^{\rm obs}_i$, which gives for each event a contribution
\begin{equation}\label{eq:margF1}
    \int \dd F^{\rm obs}_i \, \mathcal{L}(F^{\rm obs}_i | F_i) \, p_{\rm pop}(F_i| \pmb{\theta}_i, \pmb{\lambda}_{\rm jet})\,.
\end{equation}
The population prior on the flux is given by the relation between the latter and the source parameters determining its shape, $p_{\rm pop}(F| \pmb{\theta}_i, \pmb{\lambda}) = \delta \big( F - F(\pmb{\theta}_i, \pmb{\lambda}_{\rm jet}) \big)$, so $F$ can be integrated out of the likelihood (see e.g. \eqref{eq:Fgrb} below for the model of $F(\pmb{\theta}_i, \pmb{\lambda}_{\rm jet})$ used in this work). Moreover, the integral in \eqref{eq:margF1} can be restricted to the domain of observable EM data (i.e. those with $F^{\rm obs}>F_{\rm th}$ for a given detection threshold $F_{\rm th}$) by writing $\mathcal{L}(F^{\rm obs}_i | F_i) = \mathcal{L}({\rm det}_{\rm EM, i}, F^{\rm obs}_i | F_i) + \mathcal{L}(\neg{\rm det}_{\rm EM, i}, F^{\rm obs}_i | F_i)$ where $\neg {\rm det}_{\rm EM, i}$ denotes a non--detection. By definition, the second term is zero while the first is $\mathcal{L}({\rm det}_{\rm EM, i}, F^{\rm obs}_i | F_i) = P({\rm det}_{\rm EM, i} |F^{\rm obs}_i ) \mathcal{L}(F^{\rm obs}_i | F_i) = \mathcal{L}(F^{\rm obs}_i | F_i)$ because for detected events $ P({\rm det}_{\rm EM, i} |F^{\rm obs}_i ) =1$. So, \eqref{eq:margF1} becomes 
\begin{equation}\label{eq:margF2}
    \int_{ F^{\rm obs}_i\geq F_{\rm th} } \dd F^{\rm obs}_i \, \mathcal{L}(F^{\rm obs}_i | F(\pmb{\theta}_i, \pmb{\lambda}_{\rm jet})) \equiv P({\rm det}^i_{\rm EM} | \pmb{\theta}_i, \pmb{\lambda}_{\rm EM}) \, ,
\end{equation}
which leads again to \eqref{h_like_supp}, and we defined $\pmb{\lambda}_{\rm EM} = \{ A_0, \Theta_{\rm c}, F_{\rm th} \}$.

\section{Extended likelihood approximant}\label{app:lik}

\subsection{Effects of detector noise}\label{app:liknoise}

We start by discussing the inclusion of the effects of detector noise in the simulation.
The GW likelihood is 
\begin{equation}\label{eq:likelihood}
    -2\,\log \mathcal{L}({\cal D}_{\rm GW} \;|\; \vb*{\theta}) \propto \left( {\cal D}_{\rm GW} -h(\vb*{\theta})  \, | \, {\cal D}_{\rm GW} -h(\vb*{\theta}) \right)  \, , 
\end{equation}
with ${\cal D}_{\rm GW}=h_0+n$ denoting the signal, $h_0 = h( \vb*{\theta}_0)$ the waveform evaluated at the true parameters $ \vb*{\theta}_0$, and $n$ the noise. The notation $( \cdot | \cdot )$ indicates the inner product in the frequency domain~\cite{Maggiore:2007ulw},  $( a | b ) = 4 {\text{Re}} \int_0^{\infty} \tilde{a}^{*}(f) \tilde{b}(f)/S_n(f)$, with $S_n(f)$ the noise one-sided power spectral density (PSD), defined by $\langle \tilde{n}^{*}(f) \tilde{n}(f^{\prime})\rangle = \delta(f-f^{\prime}) S_n(f)/2$. 
The usual FIM approximation consists in expanding the likelihood at linear order around the point $h_0$, i.e.
$ h(\vb*{\theta}) \approx h_0 + h_i \delta {{\theta}}^{i}$ (with $h_i \equiv {\partial_{{\theta}_i}h}|_{\vb*{\theta}_0} $ and $\delta \theta^{i}=\theta^i - \theta_0^i$), obtaining\,\footnote{Here and in the following, we drop parameter--independent normalization factors.}
\begin{equation}\label{eq:LSAlikelihoodFisher}
    -2\log \mathcal{L}({\cal D}_{\rm GW} \,|\, \vb*{\theta}) \propto  \delta{{\theta}}^{i}\delta{{\theta}}^{j}  \left( h_i \, | \,  h_j\right) - 2  \delta{{\theta}}^{i} \left( n  | \,  h_i \right) \, .
\end{equation}
The covariance ${(h_i \, | \,  h_j)}^{-1}$ obtained this way does not encode the effect of noise fluctuations, as it is computed at the point $h_0$.
This is not fully consistent with the selection cut applied to obtain the GW catalog, which, as discussed previously, should depend only on the observed data, including noise fluctuations~\cite{Essick:2023upv}.\footnote{In practice the impact of ignoring this effect on the results of simulations depends on the specific problem, sensitivity, and catalog size, and should be evaluated case-by-case~\cite{Essick:2023upv}.}

We include the effect of detector noise as follows. We restrict for clarity to the case of the FIM, neglecting higher-order corrections, but the argument does not depend on this assumption. For each simulated event with true parameters $\vb*{\theta}_0$, and for each GW detector in the network under consideration, we generate an explicit realization of the noise from the PSD in the frequency domain.

For events that pass the selection cut on the observed SNR $\rho_{\rm obs}$, computed on the specific noise realization, we then find the point $\vb*{\hat \theta}$ that minimizes the negative log-likelihood in \eqref{eq:likelihood} via numerical minimization.\footnote{More specifically, after drawing a specific noise realization over a linear frequency grid, we perform a minimization helped by the fact that we know $\vb*{\theta}_0$, so the initialization helps the minimizer to converge quickly, and we incorporate physical prior ranges for all parameters.} Then, we consider the expansion of the likelihood around the ML point $\vb*{\hat \theta}$, i.e. $ h(\vb*{\theta}) \approx h(\vb*{\hat \theta}) +\hat h_i \delta \hat{\theta}^{i}$ with $\delta \hat{\theta}^{i}=\theta^i - \hat{\theta}^i $.
The notation $\hat h_i$ indicates the partial derivative evaluated at the point $\vb*{\hat \theta}$: $\hat{h}_i \equiv {\partial_{{\theta}_i}h}|_{\vb*{\hat \theta}} $. We also denote $h(\vb*{\hat \theta}) = \hat h$.

At first order we get 
\begin{equation}\label{eq:LSAlikelihoodFisherML}
    \begin{split}
        -2\log \mathcal{L}({\cal D}_{\rm GW} \,|\, \vb*{\theta}) \propto &- \delta\hat{{\theta}}^{i}\delta\hat{{\theta}}^{j}  \left(\hat h_i \, | \, \hat h_j\right) \\
        &+ 2  \delta\hat{{\theta}}^{i} \left(h_0 + n - \hat h  \, | \, \hat h_i\right) \,.
    \end{split}
\end{equation}
The first term is the usual expansion with a Fisher matrix computed around the ML point, while the second differs from \eqref{eq:LSAlikelihoodFisher} because we are expanding the likelihood around $\vb*{\hat \theta}$ and not around ${\vb*{\theta}}_0$. 
By definition, the point $\vb*{\hat \theta}$ is such that (neglecting overall normalization factors that are not relevant) 
$-2\log \mathcal{L}({\cal D}_{\rm GW} \,|\, \vb*{\hat \theta}) \approx 0$, so we must have ${\cal D}_{\rm GW} = h_0 +n \approx h(\vb*{\hat \theta}) = \hat h$, i.e the second term vanishes.
In conclusion, after finding the ML point, in the FIM approximation we can just approximate the likelihood as a multivariate Gaussian centered around $\vb*{\hat \theta}$ and with covariance $(\hat h_i \, | \, \hat h_j)^{-1}$. 

\subsection{Explicit form of the likelihood approximant}
Here we present some expressions of the extended likelihood approximant, which is exact in some subsets of parameters $\pmb{\beta}$;
while $\pmb{\beta}=\{d_L, \iota \}$ is the focus of the present work, it is also worth considering other cases.

To begin with, we consider the case $\pmb{\beta}=\{d_L\}$. The waveform parameters are then split as $\pmb{\theta} \equiv \{ d_L, \bar{\pmb{\theta}} \}$ and the waveform can be expanded around the ML point as
\begin{equation}
    h(d_L\,,\bar{\pmb{\theta}})\simeq h(d_L\,,\hat{\bar{\pmb{\theta}}}) +(\bar{\theta}^i - \hat{\bar{\theta}}^i) \cdot \partial_{\bar{\theta}_i}h(d_L\,,\bar{\pmb{\theta}})\big|_{\bar{\pmb{\theta}}=\hat{\bar{\pmb{\theta}}}}\,.
\end{equation}
The approximant obtained with this expansion will encode the effect of noise fluctuations in its width.
Replacing this expansion into \eqref{eq:likelihood} leads to
\begin{equation}\label{FdLLomng}
    \begin{split}
        -2\,\log{\cal L}( {\cal D}_{\rm GW} | d_L \,, &\bar{\pmb{\theta}}) \propto \left(\frac{ \hat{d}_{L}}{d_L}\right)^2 \Bigg[\left(\delta d_L\right)^2 \Gamma_{d_L d_L} \\
        &\ +2\delta\bar{\theta}_i\  \delta d_L \Gamma_{i d_L}+\delta\bar{\theta}_i\Gamma_{ij}\delta\bar{\theta}_j\Bigg]\,,
    \end{split}
\end{equation}
with
\begin{equation}
    \delta d_L\equiv d_L-\hat{d}_L\,,\quad \delta\bar{\theta}_i= \bar{\theta}^i - \hat{\bar{\theta}}^i\,.    
\end{equation}
The prefactor $(\hat{d}_{L}/d_L)^2$ naturally suppresses the likelihood for small values of $d_L$, thus making it unnecessary to enforce the prior $d_L>0$. This version of the likelihood could be useful to treat nearby GW sources, where the Taylor expansion in $d_L$ is less likely to be accurate.

We next consider the case $\pmb{\beta}=\{d_L\,,\iota\}$, which has been used to produce the results of this paper.
We split the waveform parameters as $\pmb{\theta} \equiv \{ d_L, \iota, \bar{\pmb{\theta}} \}$, and linearly expand the likelihood in \eqref{eq:likelihood} only in the subspace of the parameters $\bar{\pmb{\theta}}$. 
This time the waveform expansion can be written as
\begin{equation}
    h(d_L\,,\iota\,,\bar{\pmb{\theta}})\simeq h(d_L\,,\iota\,,\hat{\bar{\pmb{\theta}}}) +(\bar{\theta}^i - \hat{\bar{\theta}}^i) \cdot \partial_{\bar{\theta}_i}h(d_L\,,\iota\,,\bar{\pmb{\theta}})\big|_{\bar{\pmb{\theta}}=\hat{\bar{\pmb{\theta}}}}\,. 
\end{equation}
To express the final result, we define the polarization--decomposed FIM elements as
\begin{equation}
    \Gamma^{ab}_{i j} \equiv \frac{(f^a_{,i}|f^b_{,j})}{{\hat{d}_L}^2} \, , \quad a,b \in\{+,\,\cross\} \, ,
\end{equation}
with $f^{a,b}$ defined by
\begin{equation}
    \begin{aligned}
        & h(\pmb{\theta})=\frac{c_+(\iota)}{d_L}f^+(\bar{\pmb{\theta}})+i\frac{c_{\cross}(\iota)}{d_L}f^{\cross}(\bar{\pmb{\theta}})\,, \\
        & c_+(\iota) = \frac{1+\cos^2 \iota}{2}\, , \quad c_{\cross}(\iota) = \cos \iota \, .
    \end{aligned}
\end{equation}
where a subscript ${}_{, i}$ denotes the partial derivative with respect to $\bar{\theta}_i$.
The explicit expression for $f^{a,b}$, which is not relevant for the discussion here, can be obtained by comparing the above equations to the full frequency--domain waveform, e.g. Eq.~(22) of~\cite{Iacovelli:2022bbs}.
We find that the result can be written as
\begin{equation}\label{FdLiotaLomng}
    \begin{split}
        &-2\,\log{\cal L}( {\cal D}_{\rm GW} | d_L \,,\iota\,, \bar{\pmb{\theta}}) \propto  \\
        &\quad \left(\frac{ \hat{d}_{L}}{d_L}\right)^2\Bigg[\tilde{\delta}d^a_L \Gamma^{ab}_{d_L d_L}\tilde{\delta}d^b_L+2\tilde{\delta}\bar{\theta}^a_i\  \tilde{\delta} d^b_L \Gamma^{ab}_{i d_L}+\tilde{\delta}\bar{\theta}^a_i\Gamma_{ij}^{ab}\tilde{\delta}\bar{\theta}^b_j\Bigg]\,,
    \end{split}
\end{equation}
with
\begin{equation}
    \begin{split}
        &\Gamma^{ab}_{i d_L} \equiv -\frac{(f^a_{,i}|f^b)}{{\hat{d}_L}^3}\,,\quad \Gamma^{ab}_{d_L d_L} \equiv \frac{(f^a|f^b)}{{\hat{d}_L}^4}\,,\\
        &\tilde{\delta}\bar{\theta}_i^{+\,,\cross}\equiv \delta\bar{\theta}_i \cdot c_{+\,,\cross}(\iota)\,, \\
        &\tilde{\delta}d_L^{+\,,\cross}\equiv \delta d_L\cdot c_{+\,,\cross}(\hat{\iota}) -\hat{d}_{L} [ c_{+\,,\cross}(\iota)-c_{+\,,\cross}(\hat{\iota})]\,. 
    \end{split}
\end{equation}

The polarization--decomposed FIM elements $\Gamma^{ab}_{d_L d_L}\,,\Gamma^{ab}_{d_L i}\,, \Gamma^{ab}_{ij}$ can be derived from a combination of the usual FIM elements, evaluated at three conveniently chosen values of $\iota$. Using in particular $\iota=\{0\,,\pi/2\,,\pi\}$, we have $\Gamma_{IJ}^{++} = 4 \Gamma_{IJ}(\iota=\pi/2)$, $\Gamma_{IJ}^{\times\times} = - \left\{-\left[\Gamma_{IJ}(\iota=0)+\Gamma_{IJ}(\iota=\pi)\right]/2 + 4\Gamma_{IJ}(\iota=\pi/2)\right\}$, $\Gamma_{IJ}^{+\times} + \Gamma_{IJ}^{\times+} = \left[\Gamma_{IJ}(\iota=0)-\Gamma_{IJ}(\iota=\pi)\right]/2$, with $I(J)$ labeling either $d_L$ or $i(j)$ and $i,j$ running over $\bar{\pmb{\theta}}$. 
While in this paper we use this strategy, we point out that a more efficient approach would be to compute the FIM for each polarization directly, in which case the computational cost would be identical to that of a single FIM.
Using the same argument as in Sec.~\ref{app:liknoise}, we find that additional corrections to \eqref{FdLiotaLomng} at first order in $ \delta\hat{{\theta}}^{i}$ cancel, so evaluating \eqref{FdLiotaLomng} around the maximum likelihood point fully encodes the effects of noise fluctuations.

As a further extension, we also report the result for the case $\pmb{\beta}=\{d_L\,,\iota\,,\psi\}$, being
$\psi$ the polarization angle. This is relevant for forecasting joint GW+GRB polarization measurements~\cite{Kole:2022cdn}. In this case, the following definitions are needed
\begin{equation}
    \Gamma^{a\alpha,b\beta}_{i j} \equiv (f^{a\alpha}_{,i}|f^{b\beta}_{,j}) \, , \quad a,b \in\{+,\,\cross\}\, , \quad \alpha,\beta \in\{c,\,s\} \, ,
\end{equation}
with
\begin{equation}
    \begin{aligned}
        & h(\pmb{\theta})=\frac{c_+(\iota)}{d_L}\left[\cos{2\psi} f^{+c}(\bar{\pmb{\theta}})+\sin{2\psi} f^{+s}(\bar{\pmb{\theta}})\right] \\
        &\qquad\ +i\frac{c_{\cross}(\iota)}{d_L}\left[\cos{2\psi} f^{\cross c}(\bar{\pmb{\theta}})+\sin{2\psi} f^{\cross s}(\bar{\pmb{\theta}})\right]\, .
    \end{aligned}
\end{equation}
The likelihood \eqref{eq:likelihood} in this approximation can be formally written in a way that recalls the previous case
\begin{equation}
    \begin{split}
        &-2\,\log{\cal L}( {\cal D}_{\rm GW} | d_L \,,\iota\,,\,\psi,\, \bar{\pmb{\theta}}) \propto \left(\frac{ \hat{d}_{L}}{d_L}\right)^2\Bigg[\tilde{\delta}d^{a\alpha}_L \Gamma^{a\alpha,b\beta}_{d_L d_L}\tilde{\delta}d^{b\beta}_L \\
        &+2\tilde{\delta}\bar{\theta}^{a\alpha}_i\  \tilde{\delta} d^{b\beta}_L \Gamma^{a\alpha,b\beta}_{i d_L}+\tilde{\delta}\bar{\theta}^{a\alpha}_i\Gamma_{ij}^{a\alpha,b\beta}\tilde{\delta}\bar{\theta}^{b\beta}_j\Bigg]\,,
    \end{split}
\end{equation}
with
\begin{equation}\label{eq:likpsi}
    \begin{split}
        &\tilde{\delta}\bar{\theta}_i^{+c}\equiv \delta\bar{\theta}_i \cdot c_{+}(\iota)\cos{2\psi} \\
        &\tilde{\delta}d_L^{+c}\equiv \delta d_L\cdot c_{+}(\hat{\iota}) \cos{2\hat{\psi}} -\hat{d}_{L} [ c_{+}(\iota)\cos{2\psi}-c_{+}(\hat{\iota})\cos{2\hat{\psi}}] \, , 
    \end{split}
\end{equation}
and all the other values of $\tilde{\delta}\bar{\theta}_i^{a\alpha}$ and $\tilde{\delta}d_L^{a\alpha}$ for the indexes $a=\{ +, \cross \}$ and $\alpha=\{ c, s \}$ are obtained by replacing $+$ with $\times$ and/or $\cos2\psi$ with $\sin2\psi$ in \eqref{eq:likpsi}.

Again, the decomposed FIM matrices can be either computed directly with an adapted code or derived from the usual FIM, evaluated at specific values of $\iota$ and $\psi$ (a practical set of values is $\iota=\{0\,,\pi/2\,,\pi\}$ and $\psi=\{0\,,\pi/8\,,\pi/4\}$).

Finally, we note that this formalism can be extended to all parameters entering the signal amplitude, in particular the sky position, which is crucial for obtaining detailed predictions on localization capabilities.

\section{Population and detector models}\label{sec:simulations}

\begin{table*}[t]
    \begin{tabular}{!{\ }l l c!{\ }}
        \toprule
        \midrule
        Parameter & \multicolumn{1}{c}{Description} & Fiducial Value  \\
        \cmidrule(l{.2em}r{.2em}){1-3}
        & \textbf{Cosmology (flat $\bm{\Lambda}$CDM) -- $\pmb{\lambda}_{\rm c}$} &  \\
        $H_0$ & Hubble constant $[\si{\kilo\meter\per\second\per\mega\parsec}]$ & 70 \\
        $\Omega_{\rm m,0}$ & Matter energy density & 0.3  \\
        \cmidrule(l{.2em}r{.2em}){1-3}
        & \textbf{Rate evolution (Madau--Dickinson) -- $\pmb{\lambda}_{\rm z}$} &  \\
        $\gamma$ & Slope at $z<z_{\rm p}$ & 1.5 \\
        $\kappa$ & Slope at $z>z_{\rm p}$ & 5 \\
        $z_{\rm p}$ & Peak redshift & 2 \\
        \cmidrule(l{.2em}r{.2em}){1-3}
        & \textbf{Mass distribution (Gaussian) -- $\pmb{\lambda}_{\rm m}$ } &  \\
        $\mu_{\rm M}$ & Mean $[\si{\Msun}]$ & 1.3  \\
        $\sigma_{\rm M}$ & Standard deviation  $[\si{\Msun}]$ & 0.09  \\
        \cmidrule(l{.2em}r{.2em}){1-3}
        & \textbf{EM detection probability  -- $\pmb{\lambda}_{\rm EM}$} &  \\
        $A_0$ & Amplitude of the flux $[\si{\erg\per\second}]$ & \num{2.7e51}  \\
        $\Theta_{\rm c}$ & Half--opening of the Gaussian jet profile  $[\rm deg]$ & 3.27 \\
        $F_{\rm th}$ & GRB detector flux threshold $ [\si{\erg\per\square\centi\meter\per\second}]$ & \num{2.7e-07}  \\
        \cmidrule(l{.2em}r{.2em}){1-3}
        & \textbf{Single event parameters  -- $\{\pmb{\theta}_{\rm i}\}_{i=1}^{N_{\rm obs}}$} &  \\
        $m_{1}^{\rm det}$ & Detector--frame primary mass $[\si{\Msun}]$ & --  \\
        $m_{2}^{\rm det}$ & Detector--frame secondary mass $[\si{\Msun}]$ & -- \\
        $d_L$ & Luminosity distance $[\rm Gpc]$ & -- \\
        $\iota$ & Inclination $[\rm deg]$ & -- \\
        \midrule
        \bottomrule
    \end{tabular}
    \caption{Summary of the population hyperparameters $\pmb{\lambda}$ with their fiducial values, and of the individual event parameters $\{\pmb{\theta}_{\rm i}\}_{i=1}^{N_{\rm obs}}$ used in the hierarchical model.}\label{tab:parameters}
\end{table*}

\subsection{Populations and detectors}

For the GW population we assume $p( m_1, m_2 | \pmb{\lambda}_{\rm m}) = \mathcal{N}(m_1 | \mu_{M}, \sigma_{M}) \, \mathcal{N}(m_2 | \mu_{M}, \sigma_{M}) $ with $\mathcal{N}(x|\mu_{M}, \sigma_{M})$ a Gaussian distribution with fiducial mean and standard deviation $\{\mu_{M,{\rm fid}}=\SI{1.3}{\Msun},\, \sigma_{M,{\rm fid}}=\SI{0.09}{\Msun}\}$~\cite{Farrow:2019xnc}.
The merger rate evolution with redshift in \eqref{ppop} is modeled as a Madau-Dickinson (MD) profile~\cite{Madau:2014bja}, $\psi(z ; \pmb{\lambda}_{\rm z}) = (1+z)^{\gamma}/\{ 1 +   [ (1+z)/(1+z_\mathrm{p})]^{\gamma + \kappa} \}$.
We use fiducial parameters $\pmb{\lambda}_{\rm z} =\{ \gamma_{\rm fid} = 1.5,\, \kappa_{\rm fid} = 5,\, z_{{\rm p},{\rm fid}} =2  \}$. The fiducial values we use are obtained by convolving a MD profile with parameters determined by the Star Formation Rate~\cite{Madau:2014bja}, with a time-delay distribution. This follows from the typical, reasonable assumption that the coalescence rate of compact objects tracks the star formation rate, but with a delay taking into account the time from the binary formation to merger. The time delay follows itself some probability distribution, that we choose to be log-flat, i.e. $P(t_d) \propto 1/t_d$, as usual in the literature following from simulations and observations (see~\cite{Regimbau:2009rk} and references therein). We assume a minimum time delay of 20 Myr~\cite{Belczynski2000wr,Regimbau:2009rk}. The profile obtained after the convolution is well fit by another MD profile with the aforementioned parameters (we refer to App. A of~\cite{Iacovelli:2022bbs} for details).
%
%
Inclinations are drawn from a distribution uniform on the sphere, i.e. flat in $\cos \iota$. Since this is the same prior used for the analysis of individual GW events, there is no need to include explicitly this term in the hierarchical likelihood since it will simplify between numerator and denominator~\cite{Mandel:2018mve}.
Finally, cosmology is assumed to be described by a flat $\Lambda$CDM model with $H_0 = \SI{70}{\kilo\meter\per\second\per\mega\parsec}$, $\Omega_{\rm m,0} = 0.3$.

We consider an observing scenario representative of the best case for the current generation of GW detectors, with a LIGO-Virgo-KAGRA-LIGO India network composed of two LIGO detectors in the US, the Virgo and KAGRA detectors in Italy and Japan, respectively, and a LIGO detector located in India, with A$+$ sensitivity.\footnote{Specifically, we use the \texttt{AplusDesign} public sensitivity curve for the three LIGO detectors, \texttt{avirgo\_O5low\_NEW} for Virgo, and \texttt{kagra\_80Mpc} for KAGRA~\cite{LVKIliving}, available at \url{https://dcc.ligo.org/LIGO-T2000012/public}.}
We assume a $100\%$ duty cycle and a SNR threshold of $\rho_{\rm obs} \geq 12$, where the ``observed'' SNR is simulated encoding the effects of detector noise.

For the EM detection model, we assume that the jet has a Gaussian-shaped profile, $E(\Theta) = E_0 \exp{-\left({\Theta^2/\Theta_{\rm c}^2}\right)/2}$, (with $ \Theta = \rm{min}\{\iota, 180^{\circ} - \iota \}$  being the viewing angle) and we model the detection probability with a threshold on the observed $\gamma$-ray flux of each event.\footnote{We assume the uncertainty in the measurement of the flux to be negligible.} This is defined as $F_{\rm GRB} = \eta \, E(\Theta)/(4\pi d_L^2 T_{90})$, with $\eta=0.1$ being the radiative efficiency~\cite{Yu:2021nvx,Zhang:2003zg}, and $T_{90}$ the duration of the period during which $90\%$ of the burst's energy is emitted, which we assume to be $\SI{2}{\second}$. 
Therefore, we define $A_0 = \eta E_0/T_{90}$, and we have explicitly
\begin{equation}\label{eq:Fgrb}
    F_{\rm GRB} = \frac{A_0}{4\pi d_L^2} \, e^{ -\nicefrac{\Theta^2}{2\Theta_{\rm c}^2} }\,.
\end{equation}
    
We assume as fiducial values $E_{0,{\rm fid}} = \SI[parse-numbers = false]{10^{52.73}}{\erg}, \; \Theta_{\rm c, {\rm fid}}=\SI{3.27}{\degree}$ ($\SI{0.057}{\radiant}$), obtained by a multimessenger analysis of GW170817~\cite{Troja:2018ruz}.
We will consider an event as detected if the flux exceeds the Fermi-GBM-like sensitivity of $F_{\rm th} = \SI{2e-7}{\erg\per\square\centi\meter\per\second}$~\cite{Meegan:2009qu}.\footnote{This model corresponds to the one used in~\cite{Howell:2018nhu,Belgacem:2019tbw}, with the slight simplification that they introduce a distribution on the peak luminosity $E_0$, while here we consider it fixed. We have checked that the scaling of $\Theta_{\rm max}$ with redshift is consistent among the two.} 
In conclusion, we define $\pmb{\lambda}_{\rm EM} = \{ A_0, \Theta_{\rm c}, F_{\rm th} \}$, and we have 
\begin{equation}
    P({\rm det}_{\rm EM} |  \Theta, d_L,  A_0,  \Theta_{\rm c}) = \Bigg\{ 
    \begin{array}{c l}
        1 & {\rm if}\ F_{\rm GRB}(\Theta, d_L, A_0,  \Theta_{\rm c}) \geq F_{\rm th} \\
        0 & {\rm otherwise}
    \end{array}
\end{equation}

\autoref{tab:parameters} summarizes the population hyperparameters and individual event parameters used in the paper.

\subsection{Mock dataset}

We simulate the GW likelihood for each joint detection by using the likelihood approximant described in Sec.~\ref{app:lik}. We use the public software \texttt{GWFAST}~\cite{Iacovelli:2022mbg,Iacovelli:2022bbs} for the evaluation of the FIM, and we extract samples from the posterior given by \eqref{FdLiotaLomng} multiplied by a prior (denoted by $\pi(\pmb{\theta}_i) \, \forall i$) flat in $\cos(\iota)$ with $\iota \in [0, \pi]$, flat in detector-frame masses, and flat in the other variables. This reproduces actual data analysis choices.
We use the affine-invariant MCMC sampler \texttt{zeus}~\cite{Karamanis:2021tsx,Karamanis:2020zss}.
We assume that the sky position is known and fixed to the host galaxy, and the GPS time of detection is known.\footnote{To limit the computational cost, we also fix the phase at coalescence, polarization angle, and tidal deformabilities. We checked that this does not lead to underestimation of the uncertainty on the masses, distance, and inclination. On the other hand, fixing the spins or neglecting them would lead to substantial underestimation of the uncertainty on the detector-frame mass.}

For the EM data, we assume a Gaussian likelihood with standard deviation $\sigma_z = 10^{-3}$. We simulate a redshift measurement by drawing, for each event that passes the selection cut in the GW and EM, an ``observed'' value from this likelihood.

In summary, this procedure yields a set of samples from the posterior $p(\pmb{\theta}_i | {\cal D}_{\rm GW}^i) \propto \mathcal{L}\big( {\cal D}_{\rm GW}^i| \pmb{\theta}_i \big) / \pi(\pmb{\theta}_i)$ for each GW event and a redshift measurement, as in data analysis of real events.

Finally, to estimate the selection effect $P({\rm det} | \pmb{\lambda})$ in \eqref{h_like_supp} we use reweighted Monte Carlo integration~\cite{Tiwari:2017ndi,Farr:2019twy}. We generate a large injection set and apply the same selection cut used for constructing the catalog for GW observations, while crucially no EM selection is applied to the injections, as the EM selection probability will be derived during inference.

\begin{figure}[t]
    \centering
    \includegraphics[width=.45\textwidth]{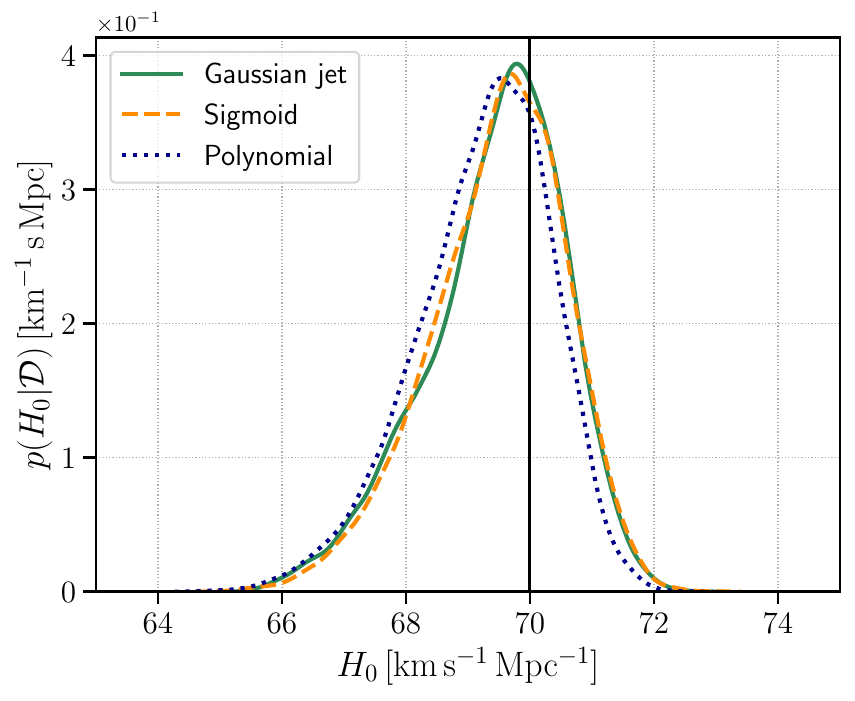}
    \caption{ Marginal posterior probability on $H_0$ with 50 multimessenger events, assuming different models for the EM detection probability (see Sec.~\ref{sec:Pmodels}).} 
    \label{fig:H0models}
\end{figure}

\subsection{Other parametrizations of \texorpdfstring{$P({\rm det}_{\rm EM} | \pmb{\lambda}_{\rm EM})$}{the electromagnetic detection probability}}\label{sec:Pmodels}

In order to ensure that the inference scheme is robust to the modeling of the EM detection probability, we check that analyzing the data (still generated with the model described in the previous section) with the assumption of a functional form for $P({\rm det}_{\rm EM} | \pmb{\lambda}_{\rm EM})$ different from the one used in the generative process still leads to an accurate determination of $H_0$. In particular, we adopt two alternative functional forms. The first is defined as
\begin{equation}\label{eq:Pdet_sigmoid}
    P_{\sigma}({\rm det}_{\rm EM} | \pmb{\lambda}_{\rm EM}) = \frac{1 - \sigma(\Theta, \Theta_{\rm M}, \sigma_{\Theta})}{1 -\sigma(0, \Theta_{\rm M}, \sigma_{ \Theta} )} \, ,
\end{equation}
with $\sigma(\Theta, \Theta_{\rm M}, \sigma_{\Theta} ) = 1/(1+e^{-\nicefrac{(\Theta - \Theta_{\rm M})}{\sigma_{\Theta})}}$ being a sigmoid function. The second is a fourth-order polynomial, 
\begin{equation}\label{eq:Pdet_poly}
    \log P_{\rm p}({\rm det}_{\rm EM} | \pmb{\lambda}_{\rm EM}) =  - a_1 \Theta - a_2 \Theta^2 - a_3 \Theta^3 - a_4 \Theta^4 \,,
\end{equation}
where the $\log$ is used to ensure positivity of $P_{p}({\rm det}_{\rm EM} | \pmb{\lambda}_{\rm EM})$ and positive priors are used on the coefficients. Both functions are normalised so that $P(\Theta=0)=1$.
Using both models, we run the inference on the largest sample used for the main results of the paper (50 events) in which case the bias, if any, should be more visible. We find that the posterior distribution on $H_0$ is not significantly affected by the specific choice of the functional form for $P({\rm det}_{\rm EM} | \pmb{\lambda}_{\rm EM})$. This is shown in \autoref{fig:H0models}.

\subsection{Hierarchical inference}

Starting from the mock data, we sample \eqref{h_like_supp} in the high-dimensional space of $\pmb{\lambda}, \{ \pmb{\theta}_i \}$ using \texttt{PYMC}~\cite{pymc2023} and the Hamiltonian Monte-Carlo-based scheme NUTS~\cite{Brooks_2011,JMLR:v15:hoffman14a}.
This choice is dictated by the necessity of accurately exploring the posterior distribution of individual GW events in the corner of the parameter space at small inclination which is preferentially selected by the EM detection probability. Using standard approaches that rely on Monte Carlo integrals over existing posterior samples~\cite{Mandel:2018mve} would either result in numerically unreliable results or require an unreasonably large amount of individual posterior samples.
In our implementation, prior to inference we instead fit a continuous interpolant of the individual GW likelihoods with a Gaussian Mixture Model over posterior samples~\cite{pedregosa2011scikit}, from which we can continuously resample at inference time.

\putbib
\end{bibunit}

\end{document}